\documentclass{amsart}[11pt]
\usepackage{mathtools}
\usepackage{hyperref}
\hypersetup{hidelinks}

\usepackage{bm, xspace}  
\usepackage{multirow}
\newcommand{\bmath}[1]{\ensuremath{\bm{#1}}\xspace}

\newcommand{\balpha}{\bmath{\alpha}}

\newcommand{\br}{\bmath{r}}

\newcommand{\bx}{\bmath{x}}

\newcommand{\bSigma}{\bmath{\Sigma} }

\usepackage{algorithm, algorithmicx}

\algnewcommand\algorithmicinput{\textbf{Input:}}
\algnewcommand\Input{\item[\algorithmicinput]}
\algnewcommand\algorithmicinitilize{\textbf{Initilize:}}
\algnewcommand\Initilize{\item[\algorithmicinitilize]}
\algnewcommand\algorithmicoutput{\textbf{Output:}}
\algnewcommand\Output{\item[\algorithmicoutput]}
\algnewcommand\algorithmicalg{\textbf{Algorithm:}}
\algnewcommand\Algorithm{\item[\algorithmicalg]}
\algnewcommand\algorithmicnote{\textbf{Note:}}
\algnewcommand\Note{\item[\algorithmicnote]}

\begin{document}
	
\title[Optimal Design of Clinical Trials]{The Optimal Design of Clinical Trials with Potential Biomarker Effects, A Novel Computational Approach}
\author{ Yitao Lu$^{1,2}$, Julie Zhou$^{1}$, Li Xing $^{3}$,  Xuekui Zhang $^{1, *}$}
\address{
$^{\text{\sf 1}}$ Department of Mathematics and Statistics, University of Victoria, Victoria, BC, V8N 1Y2, Canada\\
$^{\text{\sf 2}}$ Department of Finance and Statistics, University of Sci. \& Tech. of China, Anhui, China, Canada\\
$^{\text{\sf 3}}$ Department of Mathematics and Statistics, University of Saskatchewan, Saskatoon, SK, S7N 5E6, Canada\\
$^{\text{\sf *}}$ Corresponding author: Xuekui@UVic.ca}
\email{Xuekui@UVic.ca}

\date{\today}     
	
\begin{abstract}
	As a future trend of healthcare, personalized medicine tailors medical treatments to individual patients. It requires to identify a subset of patients with the best response to treatment. The subset can be defined by a biomarker (e.g. expression of a gene) and its cutoff value. Topics on subset identification have received massive attention. There are over 2 million hits by keyword searches on Google Scholar. However, how to properly incorporate the identified subsets/biomarkers to design clinical trials is not trivial and rarely discussed in the literature, which leads to a gap between research results and real-world drug development.
    
    To fill in this gap, we formulate the problem of clinical trial design into an optimization problem involving high-dimensional integration, and propose a novel computational solution based on Monte-Carlo and smoothing methods.  Our method utilizes the modern techniques of General-Purpose computing on Graphics Processing Units for large-scale parallel computing. Compared to the standard method in three-dimensional problems, our approach is more accurate and 133 times faster. This advantage increases when dimensionality increases.  Our method is scalable to higher-dimensional problems since the precision bound is a finite number not affected by dimensionality.
                
     Our software will be available on GitHub and CRAN, which can be applied to guide the design of clinical trials to incorporate the biomarker better. Although our research is motivated by the design of clinical trials, the method can be used widely to solve other optimization problems involving high-dimensional integration. 

\end{abstract}
\maketitle	
\pagestyle{plain}
	
\section{Introduction} \label{sec1}
Personalized medicine is the future trend of healthcare, which tailors medical treatments to individual patients. When treatments are given to the correct subset of patients who have positive responses, patients can receive the most suitable treatment according to their personal characteristics, and pharmaceutical companies can increase the chance of success in their clinical trials. The subset of patients who respond to a drug is usually defined by a predictive biomarker (e.g. expression level of a gene or a protein) and its threshold value. Biomarkers are typically molecular (e.g. expression level of a gene) related to increased benefit (or toxicity) from a particular drug \cite{ref1}.
For example, when HER2 expression level exceeds a threshold (i.e. over-expressed), patients respond well to multiple cancer drugs, including Trastuzumab \cite{ref3}.  Clinical trials with patient selection based on the expression level of HER2 typically show a positive effect in approximately 20\% of metastatic breast cancer patients \cite{ref3, ref4}. Similarly, the Pembrolizumab is a new immune-oncology drug, which is only prescribed to cancer patients with high PD-L1 levels \cite{ref0}.  On the other hand, here is an example of failed drug development due to its ignorance of predictive biomarkers. In 2003, the US Food \& Drug Administration (FDA) gave accelerated approval to Gefitinib for lung cancer patients who failed with platinum-based chemotherapy. This approval allows the drug to be sold to patients before its Phase \uppercase\expandafter{\romannumeral3} trial, but its formal approval still requests upon a successful Phase \uppercase\expandafter{\romannumeral3} trial. However, in 2005, FDA withdrew the approval because of lack of evidence of drug efficacy in the subsequent Phase \uppercase\expandafter{\romannumeral3} trial. The post-trial analysis found that only patients with gain-of-function mutations in the ATP-binding domain of the receptor respond to Gefitinib \cite{ref5}. In 2015, the FDA granted final approval Gefitinib to patients with EFGR gene mutations as the biomarker. In summary, using biomarkers to select subpopulations for clinical trials not only can increase the chance of success but also decrease the costs (time and money) in drug development. More importantly, we can save lives by forcing non-responders to use alternative treatments. 
	
In the community of clinical and biology research, `subset identification' received massive attention. As of May 01, 2020, there are about $2,140,000$ hits by searching this keyword on Google Scholar. However, there is a gap between the research of subset/biomarker identification and its application in real-world drug development. To fill in this gap, we need to pay extra attention as it is challenging to incorporate the identified subsets/biomarkers into the design of Phase \uppercase\expandafter{\romannumeral3} clinical trials. In drug development, Phase \uppercase\expandafter{\romannumeral3} clinical trials are the most important trials, since their purposes are to confirm the drug efficacy and also get approval to the market from regulatory agencies (e.g. FDA). The challenge of using subsets in real-world drug development is that, before conducting Phase  \uppercase\expandafter{\romannumeral3} studies, we hardly ever have enough data to confirm the correctness of identified subsets or biomarkers. Hence, it is hard to decide whether or not to use them in trial design, because there is a considerable risk associated with either decision. On the other hand, it is also not practical to use the post-trial data to confirm the subsets or biomarkers, because regulatory agencies can not directly accept the post-study-identified biomarkers or subsets. FDA requests an additional Phase  \uppercase\expandafter{\romannumeral3} trial to confirm the results, which leads to a substantial financial loss of multi-million dollars and the delay of multiple years in highly competitive competitions of drug development. Such significant losses are not acceptable for most companies.

Before conducting Phase  \uppercase\expandafter{\romannumeral3} trials, most often, we know a potential predictive biomarker (e.g. the expression level of a gene or a protein) but need to figure out its threshold value to identify the subset. Trying multiple threshold values of a biomarker leads to considering drug efficacy in multiple subsets of patients and the entire population. At the design stage, we can seldom know which subset (or entire population) is the true responder set.  The current standard approach of the pharmaceutical industry is to migrate the risks by testing drug efficacy both on the entire population and all candidate subsets. When multiple tests are conducted in a single clinical trial, regulatory agencies require controlling the Family-Wised Error Rates (FWER) defined as the probability of false positive in any of the conducted tests. To control FWER, Freidlin et al. \cite{ref7} and Jiang et al. \cite{ref6} proposed designs for alpha-allocation, which enforce the sum of the p-value thresholds of all tests being equal to $0.05$.  As a generalization of Bonferroni correction, Alpha-allocation keeps the assumption that all tests are independent. However, these tests are not independent since they are conducted on the `nested' populations. Therefore, an adjusted method is proposed in Chen et al. \cite{ref2} to incorporate such dependency. And it allows controlling FWER level at $0.05$, while the sum of all p-value thresholds exceeds $0.05$. The authors formulated the problem of the clinical trial design into a constrained optimization problem, which maximizes the study power with the constraint of controlling FWER no more than $0.05$. They provided a computational solution to find the best p-value thresholds for the tests on two nested populations (`one' subset and the entire population). They proposed the optimization problem for two-subset (i.e. three nested population) design problem, but did not provide a computational solution and application examples. 
In real-world applications, the major challenge in multi-subsets design is the heavy computation load in solving such optimization problems involving higher dimensional integrations. Chen et al. \cite{ref8} provided a computationally feasible strategy to solve multi-subsets clinical design problems. However, this approach over-simplified the optimization problem by ignoring uncertainty in the design parameters (i.e. replacing the prior distributions by fixed values). 

In this paper, we consider the more realistic optimization problems proposed in Chen et al. \cite{ref2} and propose a novel computational approach to find the optimal solution for multi-subsets problems in a reasonable amount of time. To solve the optimization problems with high-dimensional integrations, we use on the Monte-Carlo and smoothing methods. Due to the popularity of deep learning, the software and hardware support in General-Purpose computing using Graphics Processing Units (GPU) have exploded recently. We design our algorithm to be compatible with GPU computing, which is $133$ times faster than a standard algorithm using CPU to solve optimization problems of  three nested populations. The effect of speed boost increases when dimensionality increases. Our method is also scalable to higher-dimensional problems since the precision bound is a finite number, which is not affected by dimensionality. 
	
In the rest of the paper, in Section~\ref{sec2}, we provide notations, introduce problems and challenges in the design of multi-subpopulation clinical trials, and propose a novel computational approach to the optimal design; in Section~\ref{sec3}, we illustrate the applications of our method to design multiple clinical trials and compare their speed and precision with the optimal designs obtained by the standard algorithm. Section 4 and 5 contain discussions and conclusions.  We provide all mathematical proofs and derivations in the Appendix.

\section{Method}\label{sec2}
\subsection{Notations and the Optimization Problem}\label{sec2.1}
	
In Phase  \uppercase\expandafter{\romannumeral3} clinical trials, the primary outcome is a time-to-event variable, such as overall survival or progression-free survival in oncology trials. The drug efficacy in the study is tested based on the fitted Cox regression model, specifically based on the Z-statistics defined by the treatment effect on the drug versus placebo. Such tests are conducted on the patients entire population and their multiple subsets simultaneously. We keep the setting used in \cite{ref2} and extend the problem to the higher dimensional spaces. Therefore, we assume the potential predictive biomarker is a continuous variable (e.g. the expression level of a gene or a protein). A threshold value of this biomarker is needed to distinguish responders and non-responders, but this value is unknown. 

We assume $(n-1)$ different threshold values will be tested, which correspond to $(n-1)$ nested sub-populations of patients. In the notation below, we order the nested populations by their sample size from large to small.  We denote the $i$-th subpopulation contains $r_i$ proportion of patients in the entire population, for $i=1, \ldots, n$ and $1=r_1>r_2>r_3 > \ldots > r_n>0$. As defined in \cite{ref2}, the total information units are denoted by $I_3$, which is  $25\%$ of the total number of events expected to be observed in the entire population, hence the information units for the $i$-th subset are $r_i I_3$. The value of $I_3$ is decided by the sample size of the clinical trial, and it is considered as fixed input in our optimization problem. We denote $\mathbf{X} \coloneqq (X_1, X_2, \ldots, X_n)$ to be the Z-statistics of treatment effects in these $n$ nested populations.
	
	We compare the Z-statistic $X_i$ against the threshold value $Z_{1-\alpha_i}$, which is the $1-\alpha_i$ quantile of standard normal distribution.  If the test is significant in any population $i \in \{1, \ldots, n\}$, the drug is considered as effective, and the largest effective subpopulation is defined as responders. The task of clinical trial design is to decide the values of the significant levels $\mathbf{\alpha} \coloneqq (\alpha_1,\alpha_2,...,\alpha_n)$, which maximize study power and keep overall type I error ( i.e. FWER) under control. 
	
	Under the null hypothesis, the drug is not effective in any of these nested populations, hence $X_i \sim N(0,1)$ for all $i$'s. In Section A.1, we derive that the joint distribution of these Z-statistics is a multivariate normal as
{\small
	\begin{align} \label{Formula1}
	\left(\begin{array}{c} X_1\\ X_2\\ X_3\\ \vdots\\ X_n \end{array}\right)
	\sim 
	N\left(
	\mathbf{\mu}=\left(\begin{array}{c} 0\\ 0\\ 0\\ \vdots\\ 0 \end{array}\right)
	, 
	\bSigma= \left(
	\begin{array}{ccccc}
	1&\sqrt{r_2}&\sqrt{r_3}&\ldots&\sqrt{r_n}\\
	\sqrt{r_2}&1&\sqrt{\frac{r_3}{r_2}}&\ldots&\sqrt{\frac{r_n}{r_2}}\\
	\sqrt{r_3}&\sqrt{\frac{r_3}{r_2}}&1&\ldots&\sqrt{\frac{r_n}{r_3}}\\
	\vdots&\vdots&\vdots&\ddots&\vdots\\
	\sqrt{r_n}&\sqrt{\frac{r_n}{r_2}}&\sqrt{\frac{r_n}{r_3}}&...&1\\
	\end{array}\right)\right)
	\end{align}
}	
	We denote $\alpha_0$ as the desired upper bound of overall type I error, whose value is often set as $\alpha_0=0.05$ for one-sided test or $\alpha_0=0.025$ for two-sided test. Since the drug is considered as effective if it is significant in any tested populations, the FWER should be the probability of making any false positive call in these $n$ tests. So the constraint of optimization is 
{\small
	\begin{equation}\label{Formula2}
	\alpha_0 = 1- \mbox{P}(X_i< Z_{1-\alpha_i}, \forall i=1,\ldots,n | \mbox{ under } H_0) =
	1-\Phi_{\bSigma }(Z_{1-\alpha_1},Z_{1-\alpha_2},...,Z_{1-\alpha_n})
	\end{equation}
}
	where $\Phi_{\bSigma}()$ is the cumulative distribution function of the multivariate normal distribution defined in Formula~(\ref{Formula1}).
	
	Under the alternative hypothesis, the drug is effective in at least one population. To calculate the power, we need to specify the drug effect in each population. We describe the drug effect using $\mathbf{\Delta} \coloneqq (\Delta_1,\Delta_2,..,\Delta_n)$, where $\Delta_i$ is the hazard reduction of the survival model  in the $i$-th population. As a trivial extension of \cite{ref16, ref2}, the joint alternative distribution of $(X_1, X_2, \ldots, X_n)$ is 
{\small
	\begin{align} \label{Formula3}
	\left(\begin{array}{c} X_1\\ X_2\\ X_3\\ \vdots\\ X_n \end{array}\right)
	\sim 
	N\left(
	\mathbf{\mu}=\left(\begin{array}{c} \sqrt{I_3}\Delta_1\\ \sqrt{r_2 I_3}\Delta_2\\ \sqrt{r_3 I_3}\Delta_3\\ \vdots\\ \sqrt{r_n I_3}\Delta_n \end{array}\right)
	, 
	\bSigma= \left(
	\begin{array}{ccccc}
	1&\sqrt{r_2}&\sqrt{r_3}&\ldots&\sqrt{r_n}\\
	\sqrt{r_2}&1&\sqrt{\frac{r_3}{r_2}}&\ldots&\sqrt{\frac{r_n}{r_2}}\\
	\sqrt{r_3}&\sqrt{\frac{r_3}{r_2}}&1&\ldots&\sqrt{\frac{r_n}{r_3}}\\
	\vdots&\vdots&\vdots&\ddots&\vdots\\
	\sqrt{r_n}&\sqrt{\frac{r_n}{r_2}}&\sqrt{\frac{r_n}{r_3}}&...&1\\
	\end{array}\right)\right)
	\end{align}
}	
	
	Investigating the actual effect size of study drug is always the primary goal of clinical trials. Hence, the exact value of $\Delta_i$ is always unknown at the design stage of clinical trials. To incorporate the uncertainty of real effect size, we use the prior joint distribution (instead of fixed values) of drug effect in all these nested populations as an input parameter of study design, whose density function is denoted as $f(\mathbf{\Delta})$. Such prior knowledge of effect sizes $\Delta_i$'s are usually obtained from the analysis of the results of early phase (i.e. Phase  \uppercase\expandafter{\romannumeral1} or  \uppercase\expandafter{\romannumeral2}) trials of the same drug or relevant literature of similar studies. 
	
	The expected power with respect to the prior distribution of $\mathbf{\Delta}$ is given below in Formula (\ref{Formula4})
{\small
	\begin{equation}\label{Formula4}
	\begin{aligned}
	P(\mathbf{\alpha}) &= E_{\mathbf{\Delta}}[1- \mbox{P}(X_i<Z_{1-\alpha_i}, \forall i = 1 \ldots, n | \mbox{ under } H_1)]\\
	&=
	1-\int\Phi_{\bSigma}(Z_{1-\alpha_1}-\sqrt{r_1 I_3}\Delta_1,\ldots,Z_{1-\alpha_n}-\sqrt{r_nI_3}\Delta_n)
	\times f(\mathbf{\Delta})d\mathbf{\Delta}
	\end{aligned}
	\end{equation}
}	
	In summary, the optimal design problem of clinical trials is to find the best values of the vector $\mathbf{\alpha}$ which can be written as a constrained optimal problem below, that maximizes the expected power $P(\mathbf{\alpha}) $ defined in Formula~(\ref{Formula4}) subjected to the constraint of Formula~(\ref{Formula2}), 
{\small
	\begin{align} \label{fopt}
	\begin{cases}
	\underset{(\alpha_1,\ldots,\alpha_n)}{\text{maximize}} \quad &1-\int\Phi_{\bSigma}(Z_{1-\alpha_1}-\sqrt{r_1 I_3}\Delta_1,\ldots,Z_{1-\alpha_n}-\sqrt{r_nI_3}\Delta_n)
	\times f(\mathbf{\Delta})d\mathbf{\Delta}\\
	\textrm{subject to} \quad & 1-\Phi_{\bSigma }(Z_{1-\alpha_1},Z_{1-\alpha_2},...,Z_{1-\alpha_n}) = \alpha_0
	\end{cases}
	\end{align}
}	
	In this problem, ($r_1,..,r_n$) are fixed-value input parameters, whose values are determined by the proportion relationship of nested subpopulations. The $I_3$ is also a fixed-value parameter whose value is determined by sample size of the entire population. The prior distribution of expected drug effect $f(\mathbf{\Delta})$ is learned from early phase studies of the same drug or literature.
	
	To solve problem (\ref{fopt}), we first re-parametrize this $n$-dimensional constrained problem into a $n-1$ dimensional unconstrained problem by solving $\alpha_n$ as a function of $(\alpha_1,\alpha_2,..,\alpha_{n-1})$ from equation~(\ref{Formula2}). Then we apply the variant of the quasi-Newton method of limited-memory Broyden--Fletcher--Goldfarb--Shanno algorithm for bound-constrained optimization(L--BFGS--B) to find the solution. This algorithm approximates the BFGS using a limited amount of computer memory \cite{ref18,ref20}.

	When $n=2$, problem (\ref{fopt}) is the same as in \cite{ref2}. The extension of problem (\ref{fopt}) from two dimensional to high dimensional is not hard. The real challenge is how to overcome the heavy computational load to find the solution for the optimization problem involving higher dimensional integrations in every iteration of the quasi-Newton method. Next, we will propose our novel computation algorithms to address this challenge.

	\subsection{Calculate the High-Dimensional Integration by Monte Carlo and GPU Computing}\label{sec2.2}
	Solving problem (\ref{fopt}) requires to evaluate a large number of expected power defined in Formula~(\ref{Formula4}) for various levels of $\mathbf{\alpha}$. Evaluation of the expected power is computing a high-dimensional integration, which is time-consuming and is the major computation challenge in solving our optimization problem. This high dimensional integration can not be further reduced by analytic and closed-form solution; hence it needs to be evaluated using a computational approach. 
	The standard approach approximates such integration by fine-grids sum. The integration region is subdivided and then the sum of the area of each small region is the estimation of Formula~ (\ref{Formula4}). There are two concerns of using standard approach in our problem.  First, it is computationally impractical for high dimensional problems due to the curse of dimensionality. For example, to approximate a four-dimensional integration using only $1000$ grids in each dimension of $\Delta_i$, we need to calculate $10^{12}$ quantities and calculate their sum. Second, using hyper-rectangle or trapezoid to approximate the area under curve in each small region may lead to systematic bias. In some publications, a fixed number is multiplied to the approximation, and the number is ad-hocly chosen, but the real bias is way more complex than that.
	
	To address these two issues, we propose a better approximation algorithm based on Monte Carlo and large-scale parallel computing using GPU as follow. The precision of our method can be further improved by utilizing smoothing technique, which will be discussed in the next subsection. 
	
	We randomly generate large number ($N_1$) of samples of $\mathbf{\Delta}^{(k)}\coloneqq (\Delta^{(k)}_{1},\Delta^{(k)}_{2},...,\Delta^{(k)}_{n}), \mbox{ for } k = 1,...,N_1$ from the prior distribution with density function $ f(\mathbf{\Delta})$, and then approximate the expected power $P(\mathbf{\alpha})$ by 
	\begin{equation}\label{Formula5}
	\begin{aligned}
	\hat{P}(\mathbf{\alpha}) \approx 1-  \frac{1}{N_1}\sum_{k=1}^{N_1} \Phi_{\bSigma}(Z_{1-\alpha_1}-\sqrt{r_1I_3} \Delta^{(k)}_1,\ldots,Z_{1-\alpha_n}-\sqrt{r_nI_3}\Delta^{(k)}_n) 
	\end{aligned}
	\end{equation}
	Note the Formula~(\ref{Formula5}) requires evaluating $N_1$ functions $\Phi_{\bSigma}()$'s, which can be processed in parallel. This type of large scale parallel computing is suitable for GPU computing. The GPU computing allows parallel process tens of thousands of `simple' jobs at a time, which is much more than that of CPU computing can handle (e.g. Tesla V100 can handle 81,920 threads at most one time). 
	
	Although calculating the CDF of a multivariate normal distribution is an easy task for CPU,  it is not implemented in GPU. Hence we need to design our algorithm by further decomposing each CDF computing task to smaller and handleable jobs by GPU computing. Since CDF is a semidefinite integral, it can be approximated as the sum of results of many simpler operations. We approximate this two-layer integration (inner-layer of CDF evaluation and outer-layer of calculating power) using a two-layer sampling algorithm utilizing GPU computing, as described below. 
	
	To evaluate the functions $\Phi_{\bSigma}()$'s, we generate a large number ($N_2$) inner-layer samples of vector $\bx^{(l)} \coloneqq (x^{(l)}_1,x^{(l)}_2,..,x^{(l)}_n), l = 1,...,N_2$ from the multivariate normal distribution given by Formula~ (\ref{Formula1}). 
	For each outer-layer sample vector $\Delta^{(k)}$, we approximate the CDF function $\Phi_{\bSigma}()$ as 
	\begin{equation}\label{Formula6}
	\begin{aligned}
	\Phi_{\bSigma}(Z_{1-\alpha_1}-\sqrt I_3\Delta^{(k)}_1,\ldots,Z_{1-\alpha_n}-\sqrt{r_nI_3}\Delta^{(k)}_n)
	\approx  \frac{1}{N_2}\sum_{l=1}^{N_2} \delta(x^{(l)}_j \leq Z_{1-\alpha_j} - \sqrt{r_j I_3}\Delta^{(k)}_j, \forall j \in \{1,\ldots,n\}) 
	\end{aligned}
	\end{equation}
	where $\delta()$ is an indicator function that takes value 1 or 0 according to inside expression being true or false, respectively. 
	
	From the results of Formulas~(\ref{Formula5}) and (\ref{Formula6}), the expected power defined in Formula~(\ref{Formula4}) can be approximated by
	\begin{equation}\label{Formula7}
	\begin{aligned}
	\hat{P}(\balpha) = 1- \frac{1}{N_1N_2}\sum_{k=1}^{N_1}\sum_{l=1}^{N_2} \delta(x^{(l)}_j \leq Z_{1-\alpha_j} - \sqrt{r_j I_3}\Delta^{(k)}_j, \forall j \in \{1,\ldots,n\})	
	\end{aligned}
	\end{equation}
	
	Formula~(\ref{Formula6}) decomposites the evaluation of CDF into the sum of a lot of comparison operations, which is handleable by GPU. These $N_1\times N_2$ comparison jobs can be handled by GPU in parallel, and then the $N_1$ sum operations can also be done in parallel by GPU computing.
	
	The sample size needed in the two-layer sampling can be determined by the desired precision of the estimated power, i.e. the bound of $\mbox{Var}(\hat{P}(\balpha))$. In Appendix A.2, we derive a bound of this variance, which is given as 
	\begin{align} \label{Formula8}
	\mbox{Var}(\hat{P}(\balpha)) \leq \frac{1}{4N_1N_2},
	\end{align}	
For example, if we set $N_1=10240, N_2=20480$ in Monte Carlo, the precision of estimated power can be bounded by $\mbox{Var}(\hat{P}(\balpha)) \leq 1.19 \times 10^{-9}$. 
	
	We summarize the procedures discussed in this subsection in Algorithm 1 and Algorithm 2. The Algorithm1 is for calculating the expected power $P(\balpha)$. The Algorithm 2 is for calculating the CDF's $\Phi_{\bSigma}(Z_{1-\alpha_1}-\sqrt{r_1I_3} \Delta^{(k)}_1,\ldots,Z_{1-\alpha_n}-\sqrt{r_nI_3}\Delta^{(k)}_n) $, which is called inside Algorithm 1. 
	
		\begin{figure}[htb]
		\centering 
		\includegraphics[width=5in]{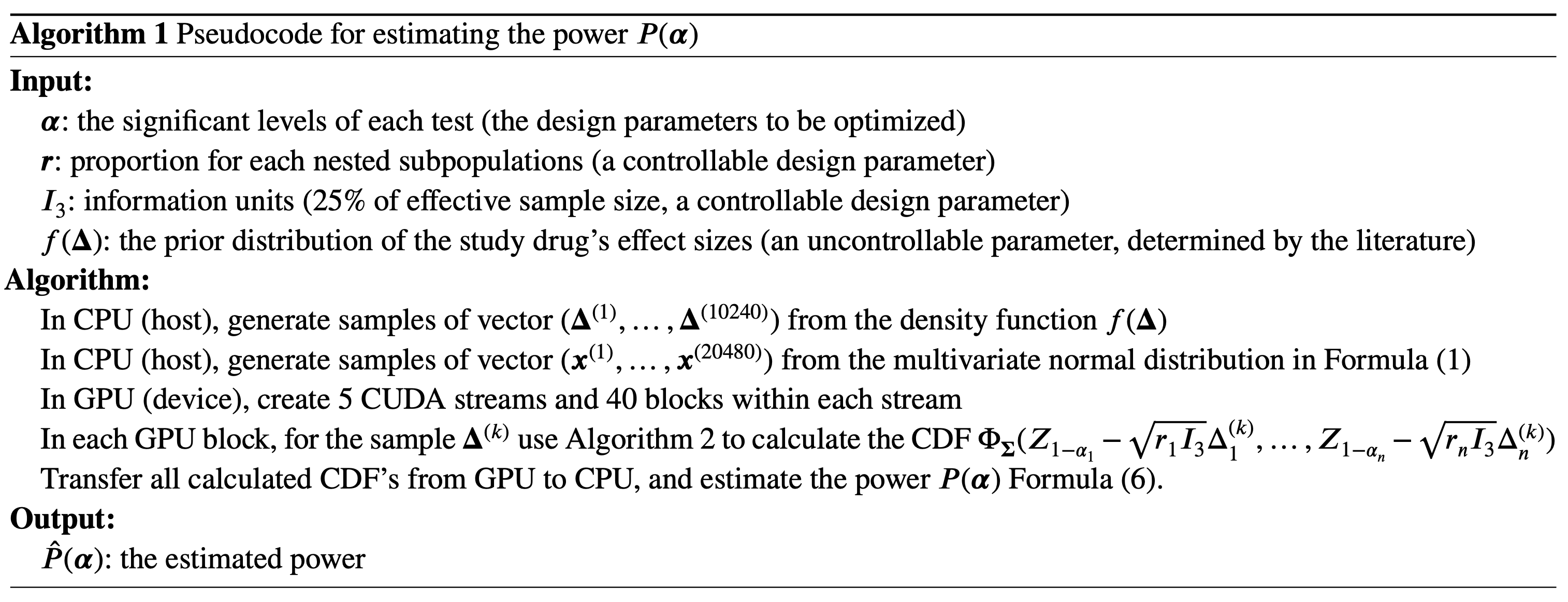}
		\label{al1}
	\end{figure}
	
	\begin{figure}[htb]
		\centering 
		\includegraphics[width=5in]{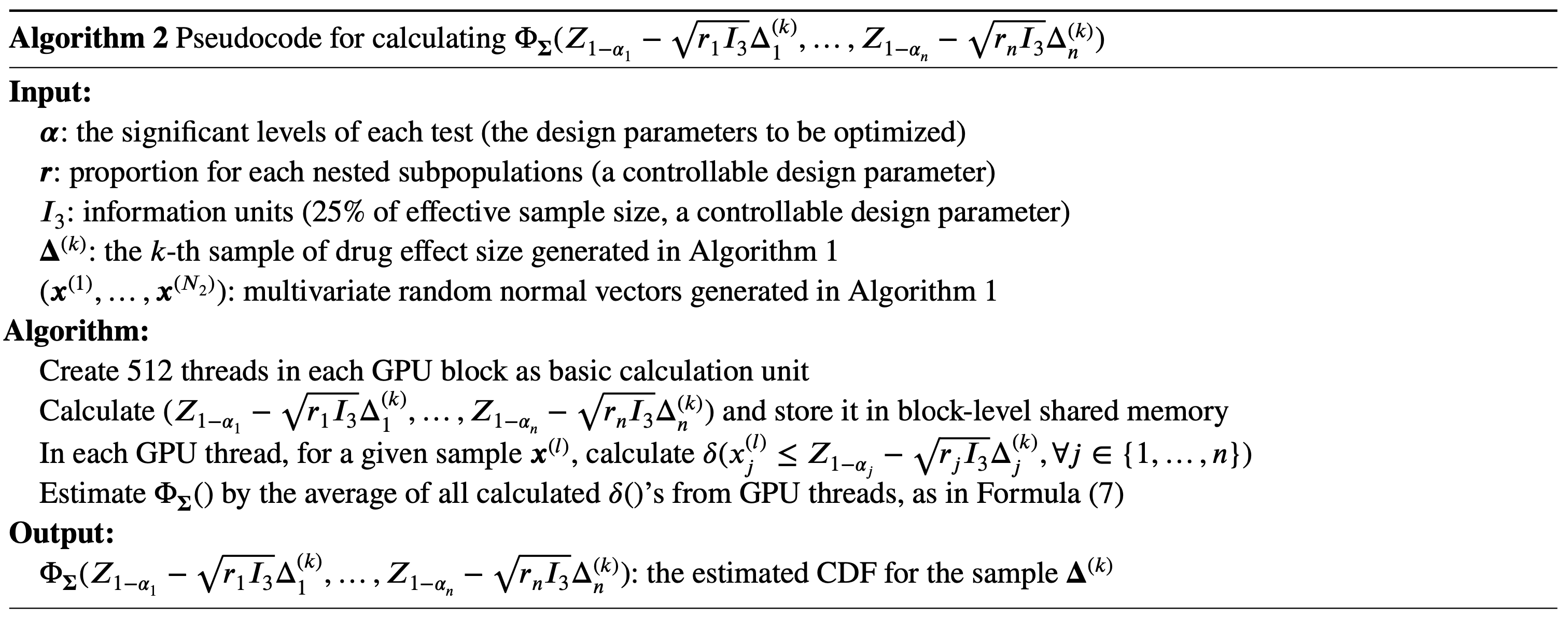}
		\label{al2}
	\end{figure}

	\subsection{Improve Accuracy by Smoothing}\label{sec2.3}
To find the optimal value of $\balpha$ that maximize the expected power, in each iteration of the Newton method, it requires calculating not only power $P(\balpha)$ for a given value of $\balpha$, but also its derivative $P'(\balpha)$ to decide the value of $\balpha$ in the next iteration. Since the derivatives cannot be calculated analytically, we have to estimate then numerically in the quasi-Newton method. The accuracy of numerical derivatives affects whether the algorithm is searching the solution in the correct direction. However small bias in estimated power $P(\balpha)$ can lead to large errors in estimated derivatives $P'(\balpha)$, which makes Newton algorithm hard to converge. To fix this problem, we propose to use a smoothing technique to improve accuracy. In Appendix A.3, we prove that $P(\balpha)$ is an (n-1)-variate infinitely differentiable function with respect to $(\alpha_1,\alpha_2,\ldots,\alpha_{n-1})$. Such smoothness property enables us to borrow information from neighboring points on the surface, and hence remove noise and improve accuracy. 

We propose to evaluate $\hat{P}(\balpha)$ at many ($N_3$) different values of $\balpha$ using Algorithm 1, and then fit a smooth surface (denoted as $\tilde{P}(\balpha)$)  to these $N_3$ points using Thin Plate Splines (TPS) \cite{ref21,ref22}. The TPS model is well known to be versatile to model any shape of high dimensional smooth surface, and it is more efficient than other functional bases like polynomial splines. We use the optimal solution on the fitted surface $\tilde{P}(\balpha)$ as approximate solution of our original optimization problem~(\ref{fopt}), since this solution can be easily and reliably found using the Newton method for three reasons.  First, we can specify good initial point to largely reduce the number of Newton iterations needed to converge. Specifically, among the $N_3$ points on the fitted surface, we select the one with the largest value of $\tilde{P}(\balpha)$ as initial point. Second, for any value of $\balpha$, we can quickly evaluate $\tilde{P}(\balpha)$ on the fitted TPS with `no' error, hence the derivatives $\tilde{P}'(\balpha)$ can be precisely estimated with low computational cost.  Third, the TPS surface enables borrowing information among neighbor points of $\hat{P}(\balpha)$'s to correct their errors via smoothing.
	
	The $N_3$ points of $\balpha$ values can be selected using the procedure as follows. For two-sided test, we set $\alpha_0=0.025$, hence $0<\alpha_i<0.025$ for $i=1, \ldots, n-1$. The solution of re-parametrized and unconstrained optimization problem should be in the $(n-1)$-dimensional hyper rectangular $(0,0.25)^{(n-1)}$. We select the $m^{(n-1)}$ points as grid spanned by equally spaced $m$ points in every dimension. Then we solve $\alpha_n$ for each spanned grid point with the constraint function in Formula (\ref{Formula2}), and identify the invalid grid points using $\alpha_n \notin [0, 0.025]$. From the valid grid points we randomly select $N_3$ points of $\balpha$. Based on our experiments, we suggest using $N_3=2000$ for 3-dimensional problems with $n=3$, and using $N_3=4000$ for 4-dimensional problems with $n=4$. But user can revise this setting according to their needs.

\subsection{Design the nested subpopulations}

 	The discussion above assumes the nested subpopulations are pre-selected, but selecting subpopulations is also an important task of clinical trial design. Selecting subpopulations of is an optimization problem respect to parameters $\br$, which could be optimized (together with $\balpha$) in problem~(\ref{fopt}). In the design of clinical trials, both the p-value thresholds $\balpha$ and the sizes of subpopulations $\br$ are controllable parameters. However, we deliberately avoid optimizing $\br$ in problem~(\ref{fopt}) for the following reason. The choice of $\balpha$ is a pure mathematics problem to maximize the study power, which is a common task for designing all clinical trials. On the other hand, the $\br$ is related to not only potential drug efficacy on final target subpopulation but also future revenue. Whether selling the drug to a slightly larger patient population while scarifying some efficacy involves business decisions, which can be affected by many factors, such as the competitors in the market and the cost-revenue-ratio etc. Hence selecting $\br$ can be guided by different utility functions in the real-world design of clinical trials. Our modeling strategy is to make selecting $\balpha$ as a blackbox component and selecting $\br$ as a component need specific users input on utility functions, which requires separate the two components in modeling.
	
	For simplicity of discussion, we use the power as the utility function in the rest of this paper to illustrate the method of selecting best values of $\br$, but users can change it to any other study-specific utility functions involving both power and other factors. To identify the best values of $\br$, we use the following procedure. First, we specify many settings of values of $\br$, calculate their corresponding optimal powers solved from problem~(\ref{fopt}). Second, we fit TPS of optimal power as functions of $\br$. At last, we find optimal solution of $\br$ on the fitted TPS using the same procedure as for the TPS $\tilde{P}(\balpha)$ discussed in the last subsection. Note, the smoothing technique can be applied for finding the optimal value of $\br$, since it is easy to prove that the function $P(\balpha)$ defined in Formula~(\ref{Formula4}) is also an indefinite differentiable function of $\br$.
	
	The complexity of our algorithm for selecting $\br$ is approximately equal to the complexity of solving optimization problem~(\ref{fopt}) multiplied by the number of settings of $\br$ used to fit TPS, since fitting TPS and finding optimal point on TPS surfaces can be done very quickly. To fit the TPS of $\br$ with decent quality, we suggest using at least a hundred of points for the three-nested-population problems, and increase the number for higher dimensional problems. Such procedure requires solving many optimization problems of (\ref{fopt}), which makes speed boosting more desired.

	Finally, the optimal design is defined by the optimal value of $\br=\hat{\br}$ together with optimal solution of $\balpha$ under the setting of $\hat{\br}$.

	\section{Applications in Design of Clinical  Trials}\label{sec3}
	
	Using examples of clinical trial designs, we illustrate the strength of our algorithm. To demonstrate the advantage of our algorithm, we compare our results and speed with a traditional algorithm. In the traditional algorithm, we use the Newton method with L-BFGS-B algorithm, but we approximate the integrations by grid sums.
	
	In these examples, we design a Phase \uppercase\expandafter{\romannumeral3} clinical trial with a potential biomarker effect. The biomarker is a continuous variable, and it has an unknown threshold value for distinguishing responder patients and nonresponders. We consider the design with two candidate threshold values and three candidate threshold values, which correspond to three nested populations and four nested-populations, respectively. To design these trials, we need to solve optimization problems with 3 and 4 dimensional integrations. For input parameters, we use the same setting as described in \cite{ref2}, but extend the structure in higher dimensional problems. We set the total number of events in the entire population of patients in Phase \uppercase\expandafter{\romannumeral3} estimated by
	\begin{equation}
		I_3 = \frac{(Z_{1-\alpha}+Z_{1-\beta})^2}{log(1-\Delta)^2}
	\end{equation}
	where $\alpha$ and $\beta$ denotes the type-\uppercase\expandafter{\romannumeral1} and type-\uppercase\expandafter{\romannumeral2} error rates, respectively, which are usually set as $\alpha=0.025$ ($5\%$ false positive rate for two sided tests) and $\beta=0.1$ ($90\%$ power), and $\Delta$ stands for the hazard reduction of the entire population. In the setting of no biomarker effect, we set the drug efficacy $\Delta $ to be a constant $0.25$. Hence, the information unit estimation $I_3$ is 127, indicating that the total number of patients' events is 508, which is typical for oncology clinical trials. In the setting of continuous biomarker effect, we set the drug efficacy $\Delta$ to be $0.2$ in the entire population. Hence, the information unit $I_3$ is 211, and the total number of patients' events is 844. 
	
	The prior distribution of effect sizes in these nested populations $(\Delta_1,\Delta_2,...,\Delta_n)$  (learned from Phase \uppercase\expandafter{\romannumeral1} and \uppercase\expandafter{\romannumeral2}) is assumed to follow a multivariate normal distribution, given by, 
{\small
	\begin{equation}
	\left(\begin{array}{c} \Delta_1\\ \Delta_2\\ \Delta_3\\ \vdots\\ \Delta_n \end{array}\right)
	\sim 
	N\left(
	\mathbf{\mu}=\left(\begin{array}{c}
	\theta_1\\
	\theta_2\\
	\theta_3\\
	\vdots\\
	\theta_n\\
	\end{array}
	\right),\mathbf{\Sigma}=\left(
	\begin{array}{ccccc}
	\sigma_1^2&\sqrt{r_2}\sigma_1\sigma_2&\sqrt{r_3}\sigma_1\sigma_3&...&\sqrt{r_n}\sigma_1\sigma_n\\
	\sqrt{r_2}\sigma_1\sigma_2&\sigma_2^2&\sqrt{r_3/r_2}\sigma_2\sigma_3&...&\sqrt{r_n/r_2}\sigma_2\sigma_n\\
	\sqrt{r_3}\sigma_1\sigma_3&\sqrt{r_3/r_2}\sigma_2\sigma_3&\sigma_3^2&...&\sqrt{r_n/r_3}\sigma_3\sigma_n\\
	\vdots&\vdots&\vdots&\ddots&\vdots\\
	\sqrt{r_n}\sigma_1\sigma_n&\sqrt{r_n/r_2}\sigma_2\sigma_n&\sqrt{r_n/r_3}\sigma_3\sigma_n&...&\sigma_n^2\\
	\end{array}\right)\right)
	\label{Formula10}
	\end{equation}
}
	where  $\theta_i = -log(1-\Delta_i^0)$,  $\Delta_i^0$ is the point estimate of hazard reduction for the $i$-th population, and we set $\sigma_i = \frac{1}{\sqrt{80*r_i/4}}$ to represent the fact that it is harder to get a precise estimation in the literature for the smaller proportion subpopulation.

In the examples of clinical trial design, we need to compare our method with the standard method. To run the standard method within reasonable computation time, we only consider the situation of $n=3$ nested populations. This leads to consider the investigation of two threshold values of biomarker. Since the sizes of subpopulations are controllable parameters in clinical trial design, we need to find the optimal setting of $\br$ that provides the best optimal power. To fit the TPS power surface of $\br$, from the grid spanned points generated by equally spaced points in $(r_2,r_3) \in (0,1)\times(0,1)$ with step size 0.05, we select all ($171$) valid pairs satisfying $1=r_1 > r_2 > r_3 > 0$.  Assuming that the drug efficacy is a linear function of the subpopulation's size, we consider three  conditions of biomarker effects: (a) No biomarker effect with drug efficacy as a constant value  $(\Delta_1^0 = \Delta_2^0 = \Delta_3^0 = 0.25)$; (b) Weak biomarker effect that is continuous as $\Delta_i^0 = 0.3 - 0.1 r_i$, which linearly increase from $0.2$ to $0.3$ when $r$ decreases from $1$ to $0$; and (c) Strong biomarker effect that is continuous as $\Delta_i^0 = 0.8 - 0.6 r_i$, which linearly increases from $0.2$ to $0.8$ as $r$ decreases from $1$ to $0$. In summary, this leads to investigating the design of clinical trials under $171\times 3= 513$ different settings. In deriving the optimal clinical designs, we use $m=50$ equally spaced grids of every dimension of $\Delta_i$ to calculate integrations in the traditional grid sum method. For our novel algorithm, we use $N_3=2000$ random selected values of $\balpha$ to fit the TPS $\tilde{P}(\balpha)$, while for each of these $N_3$ points, we calculate the integration using Monte Carlo setting of $N_1=10240, \; N_2=20480$.

\begin{figure}[b]
	\centering  
	\includegraphics[width=5in]{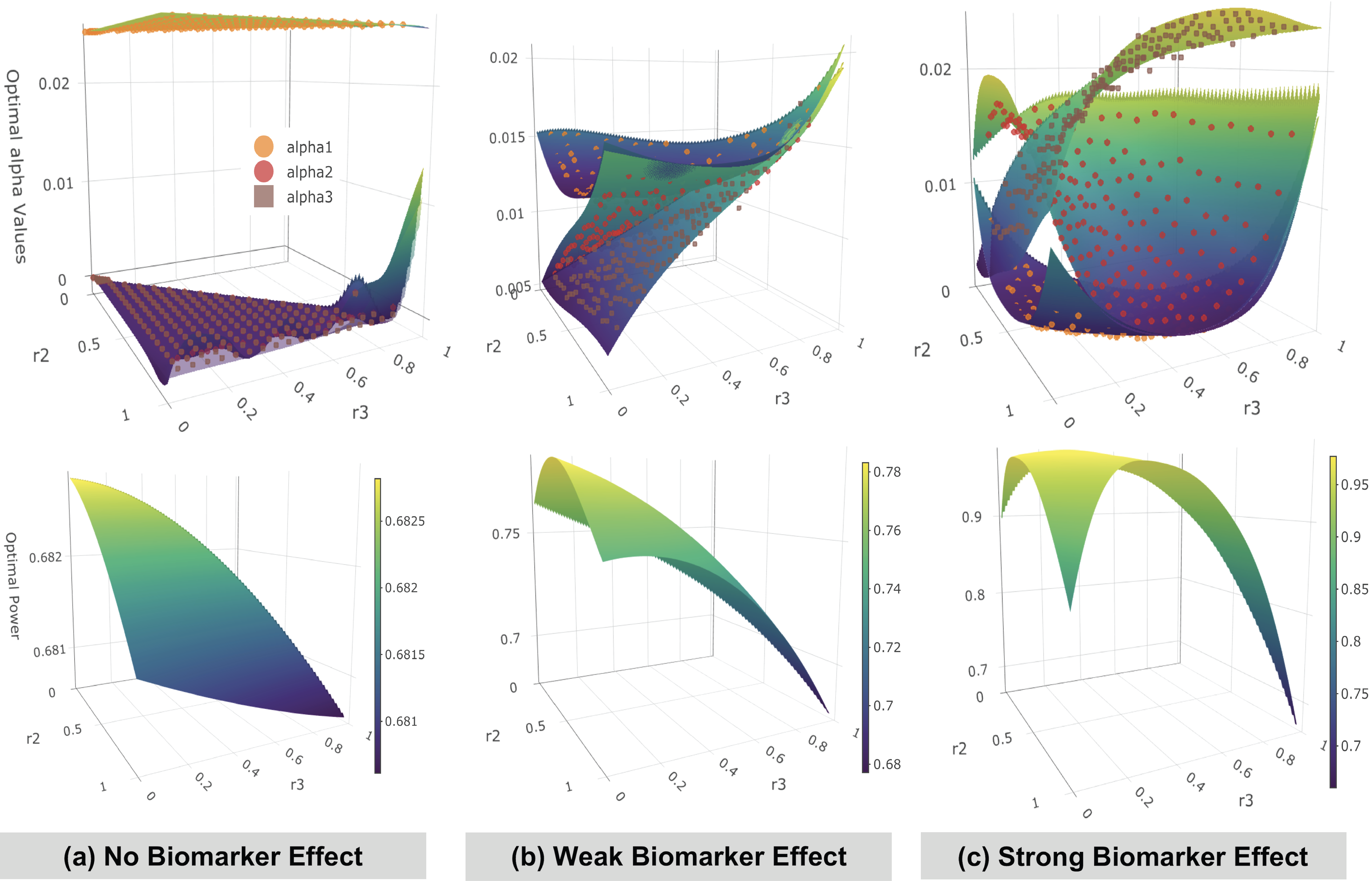}		
%
%
		\caption{{Three dimensional optimal design for (a) no biomarker effect, (b) weak biomarker effect , and (c) strong biomarker effect. The rotatable 3D plots of  these subfigures are available, whose link is given in Appendix A.4. Each column is generated from the solutions of $171$ optimization problems~(\ref{fopt}) using different settings of $r_2$ and $r_3$. The top row shows the optimal $\balpha$ values, and the bottom row shows their corresponding optimal powers. The optimal setting of $\br$ can be found on the peak of power surface in the bottom row, which guide us to define the subsets of patients.}}
		\label{Fig.1}
	\end{figure}

Figure~\ref{Fig.1} shows the optimal designs in various settings of $(r_2, r_3)$. The top row shows the optimal $\balpha$ values, while the bottom row shows their corresponding optimal powers. From the bottom row, we can learn the optimal setting of $(r_2, r_3)$ which achieve the maximum power. The rotatable 3D plots of  these subfigures are available, whose link is given in Appendix A.4.  
We find two common patterns of relationships among optimal values of $\alpha_1$, $\alpha_2$, and $\alpha_3$. First, when $r_2$ increases to be close to $1$, $\alpha_2$ increases sharply to reach the value of $\alpha_1$. We find similar relationship between the $\alpha_3$ and $r_3$. We explain this phenomenon as follows: the subpopulations become identical to the entire population when $r_2$ and $r_3$ approaching $1$. Hence, the three optimal $\alpha$'s should become identical as well. Second, the sum $\alpha_1+\alpha_2+\alpha_3$ of our optimal design can be larger than the desired FWER threshold ($0.025$ for two-sided tests), which is as expected. The alpha-splitting approaches assume all tests are independents, and have to enforce $\alpha_1+\alpha_2+\alpha_3=0.025$. In contrast, we assume the tests conducted in nested populations are related as modeled in Formula~(\ref{Formula1}) and hence allow more powerful designs as $\alpha_1+\alpha_2+\alpha_3 > 0.025$ while keeping FWER controlled at $0.025$.
	
      Figure~\ref{Fig.1}(a) shows the results of no biomarker effect conditions. The best power is $0.6847$, which is achieved at the setting of ($r_1=1, \; r_2=r_3=0$ ). This suggests only consider the entire population without subsets, which reduces our optimization problem into one dimension. In this optimal $\br$ setting, the corresponding optimal p-value thresholds are $(\alpha_1=0.025, \; \alpha_2=0, \; \alpha_3=0)$. Then $r_2$, $r_3$ are not approaching $1$ (i.e. subsets are not identical to the entire population), the optimal $\balpha$ always allocates large $\alpha_1$ to the entire population, while $\alpha_2$ and $\alpha_3$ are nearly equal to 0. This decision almost ignored all subsets, which is reasonable since the drug efficacy is a constant and the entire population has a larger sample size than the subsets.   
Figure~\ref{Fig.1}(b) shows the results of weak biomarker effect conditions. The largest optimal power is $0.783$, achieved at $(r_1=1, \; r_2=0.365, \; r_3=0)$. This suggests only consider one subpopulation instead of two, which reduces our optimization problem into two dimensions. In this optimal $\br$ setting, the optimal p-value thresholds are $(\alpha_1=0.0163, \; \alpha_2=0.0107, \; \alpha_3=0)$. The subpopulation comprises $36.5\%$ patient population. Figure~\ref{Fig.1}(c) shows the results of strong biomarker effect conditions. The largest optimal power is $0.0.977$, achieved at $(r_1=1, \; r_2=0.446, \; r_3=0.168)$, which corresponds to the optimal p-value thresholds $(\alpha_1=0.00194, \; \alpha_2=0.0135,  \; \alpha_3=0.0133)$. This suggests considering two subpopulations, which comprise $44.6\%$  and $16.8\%$ patient population, respectively. Note the stronger the biomarker effect, the larger range that drug efficacy can change, and the more subsets suggested by our optimal design. Such a decision trend agrees with our intuition.

	

\begin{figure}[htb]
		\centering 
		\includegraphics[width=4.5in]{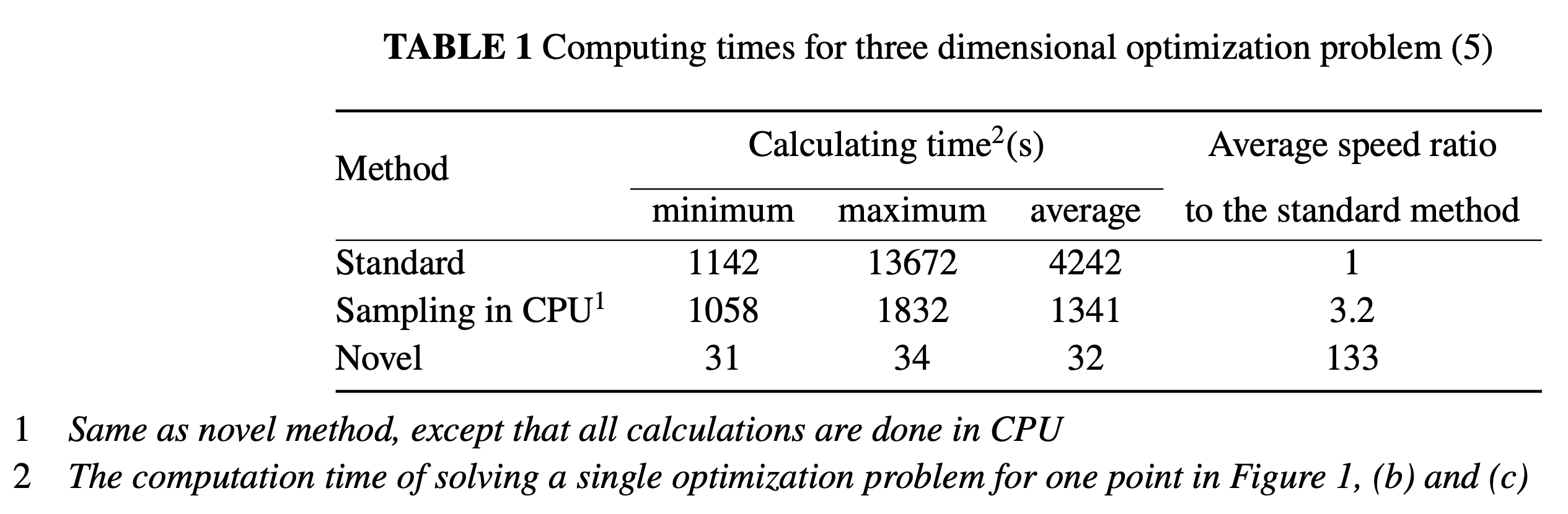}
		\label{Table1}
	\end{figure}

Table~\ref{Table1} shows the summary of computation times needed to solve a single optimization problem using each method. This summary is based on all $513$ optimization problems solved for generating Figure~\ref{Fig.1}. The speed of our novel method (using Nvidia Tesla V100 GPU on Compute Canada server Beluga) is 133 times faster than the traditional method (using the CPU on the same server). Without the help of the GPU algorithm the speed Monte Carlo approach is only 3.2 times faster than the standard grid sum approach, which illustrates the importance of utilizing GPU computing. Note the standard method uses at most $13672$ seconds to solve a single optimization problem, which seems to be acceptable. However, in the application of clinical trial design, we always need to understand the relationship of power to the settings of $r_2$ and $r_3$, which requires solving a large number of optimization problems. Solving all optimization problems used to generate Figure~\ref{Fig.1} requires $4.5$ hours to run our algorithm and over $600$ hours to run the standard grid sum algorithm. The time required could be much more if we consider the optimization problems of more than 3 dimensions. Hence, the speed boost offered by our algorithm is critical in real applications. We will discuss this in more detail in Section 4. 

%
%
%

\begin{figure}[htb]
		\centering 
		\includegraphics[width=5in]{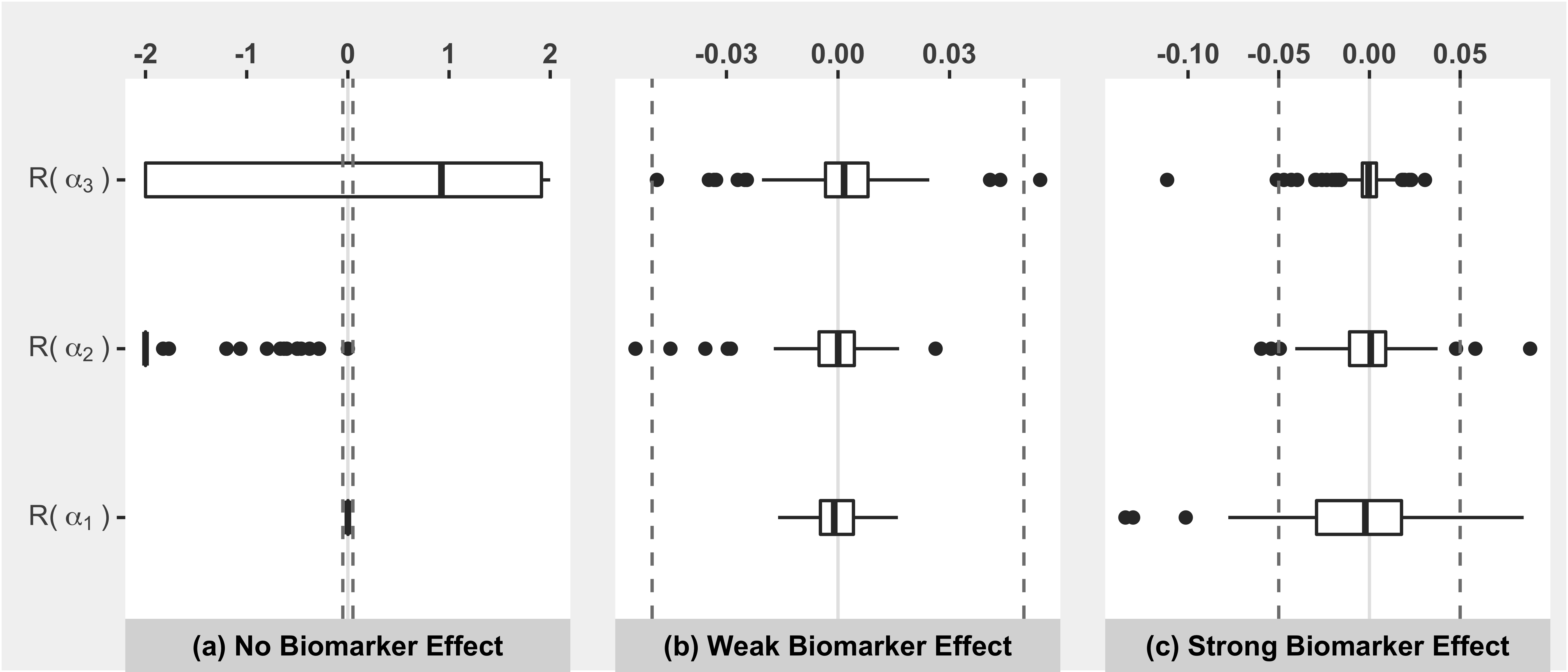}

		\caption{Boxplot of relative differences between the two methods. The relative difference is defined as $R(\alpha_i)=\frac{(\alpha_i^{(s)}-\alpha_i^{(n)})}{(\alpha^{(s)}+\alpha_i^{(n)})/2}$ for $i=1,2,3$. Where $\alpha_i^{(s)}$ and $\alpha_i^{(n)}$ denotes optimal solution of $\alpha_i$ obtained by the standard method and the novel method, respectively. Except for no biomarker effect conditions, the relative errors are well bounded by the two vertical lines at $\pm 5\%$, which means their relative difference is small.  In addition, the pairwise Wilcoxon test shows no significant difference in two methods, except for $\alpha_1$ and $\alpha_2$ in no biomarker effect conditions. Interpreting the results of no biomarker conditions are more complicated and are discussed in the main text of this paper.}
		\label{Fig.2}
	\end{figure}

To compare the difference of optimal $\balpha$ obtained from our method and standard method, we calculate their relative difference defined as 
$R(\alpha_i)=\frac{(\alpha_i^{(s)}-\alpha_i^{(n)})}{(\alpha_i^{(s)}+\alpha_i^{(n)})/2}$ for $i=1,2,3$. Where $\alpha_i^{(s)}$ and $\alpha_i^{(n)}$ denotes optimal solution of $\alpha_i$ obtained by the standard method and  the novel method, respectively. Figure~\ref{Fig.2} shows the relative difference for all alpha values estimated for the $513$ optimization problems. Except for no biomarker conditions, most relative differences are well bounded by $\pm5\%$. The optimal $\balpha$ obtained from two methods are not only similar in magnitude, but also statistically (with all non-significant p-values from pairwise Wilcoxon tests). 

In no biomarker effect conditions, the solutions of $\alpha_2$ and $\alpha_3$ are very close to $0$. Hence the relative difference should be interpreted differently. In solutions of our method, most values of $\alpha_3$ are less than $6\times10^{-6}$, hence relative difference bounded by 2 (as shown in boxplot) means both methods suggest almost ignoring the smallest subset. In addition, the Wilcoxon test p value for $\alpha_3$ is non-significant ($p=0.6$). The relative difference of $\alpha_2$ is always positive (as seen in Figure~\ref{Fig.2} ), and the relative difference of $\alpha_1$ is always negative (not enough resolution to show in Figure~\ref{Fig.2} since the absolute value of optimal $\alpha_1$ is big). Such a strong trend indicates that, compared with the standard method, our method consistently suggests spending more $\alpha$-allocation on the entire population. This a wise suggestion, since in no biomarker condition, we should only test drug efficacy on the entire population to achieve better power.

\begin{figure}[h]
		\centering
		\includegraphics[width=5in]{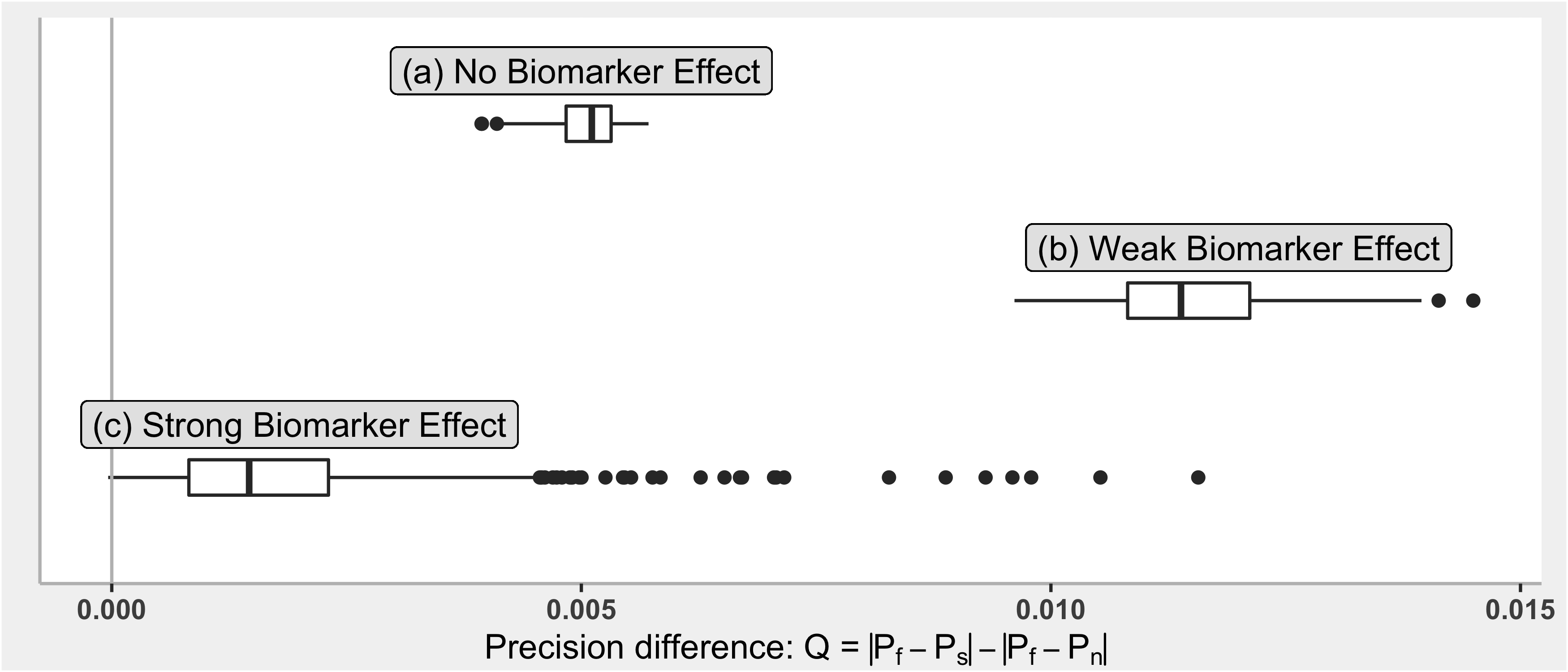}
		\caption{Boxplot of precision difference statistics $Q$, defined as $Q=|P_s-P_f|-|P_n-P_f|$. The $P_f$, $P_n$ and $P_s$ denote the optimal power obtained by "gold standard", novel, and standard method, respectively.  Positive Q values indicate that our method is more accurate than the standard method. Each box in this figure is made from the solutions of $171$ optimization problems in a biomarker effect condition. Almost all data points are above the horizontal line of $Q=0$, which shows that our novel method consistently outperforms the standard method.  }
		\label{Fig.3}
	\end{figure}

To compare the precision of estimated optimal power between our novel algorithm and the standard algorithm, we need to compare them against the `gold standard'. For each optimal setting of $\balpha$ values in the left hand side panel of Figure~\ref{Fig.1}, we calculate the expected power using a fine-grid sum approach with $m=500$ grid points for each dimension. The values calculated using fine grid sum approach should be more precise than both standard and novel method discussed above, so we use them as `gold standard' in the precision comparison. We denote the estimated power using the standard method, our novel method, and the fine grid `gold standard', and denoted as $P_s$,$P_n$, and $P_f$, respectively. For each optimization problem we solved for Figure~\ref{Fig.1}, we calculate precision difference $Q=|P_s-P_f|-|P_n-P_f|$. Positive value of statistics $Q$ indicates the solution of the novel method is closer to the truth. Figure~\ref{Fig.3} shows all $513$ Q statistics obtained from the optimization problems solved for generating Figure~\ref{Fig.1}. There is only one Q statistic slightly below the horizontal line of $Q=0$, which indicates that our method consistently provides more precise solutions than the traditional method. Wilcoxon test p-values corresponding to all three boxes/conditions are less than $2.2*10^{-16}$. We also find the performance difference between our method and the standard method is not always similar for different kinds of optimization problems. For example, the weak biomarker effect problems have much more difference than strong biomarker effect problems.
	
	\section{Discussion}\label{sec4}

       To generate Figure~\ref{Fig.1}, we need to solve $513$ optimization problems, which takes over $600$ hours to run the standard method, and $4.5$ hours to run our novel method. This illustrates the importance of speed-boosting in real-world applications. In higher dimensional design, we can speed up more significantly. Based on our limited amount of experiments in the situation of $n=4$ nested populations (results not shown), our method is over $3000$ times faster than the standard method. The following are the details about expected speed-change in $n=4$ dimensional problems. Compared with $n=3$ dimensional problems, the standard method need $70$ times more time to solve an optimization problem, because of the one more dimension of sum with grid size $m=50$ and other related calculations. In comparison, our novel method is about $3$ times slower because of the doubled value of $N_3$ and other related calculations. We use $N_3=4000$ to fit $4$-dimensional TPS of $\tilde{P}(\balpha)$, since higher dimensional surface needs more points to achieve a similar level of good fit.

We are not the first algorithm using GPU to evaluate integrals. For example, VEGAS \cite{ref11} is one of the most popular integration packages in Python. To improve accuracy without much time consumption, the algorithms in VEGAS are combined with an adaptive multi-channel sampling method \cite{ref12}.  BASES  \cite{ref13} is another popular algorithm. It deals with the integration of singular functions. Typical computer clusters can take months to perform the computations that calculate high dimensional integration, but GPU accelerated solutions can massively speed up these computations \cite{ref14, ref15}. However, these methods are mostly developed for general purpose integration. General methods are not efficiently designed to utilize properties of our specific problem, and some methods (such as ZMCIntegral\cite{ref11}) cannot handle our complex integral functions in our problem. This motivated us to implement our integral functions. In addition, the most  contribution of our method is incorporating a smoothing technique to improve precision and solve the converging issue of the Newton methods, which is not proposed in previous studies.

	
	The bound of variance Formula~(\ref{Formula8}) leads to many advantages of sampling-based methods over the standard sum-of-grid-value method. First, because the upper bound of estimation variance is a finite number regardless of dimensionality, our method is scalable to obtain the solution of higher dimensional problems with reasonable precision and reasonable time. In contrast, to keep the same precision level, the complexity of the `sum of grid values approximation' method increases exponentially with dimension.  Second, our method can reach any desired precision level by increasing the number of samples (i.e. $N_1$ and $N_2$) accordingly, whereas the sum-of-grid approach introduce a systematic error by approximating area under a curve using trapezoid shape. 
	

\section{Conclusion}\label{sec5}
In this work, we proposed a novel computational solution for optimization problems involving high dimensional integration, and applied it to the design of clinical trials. Our algorithm utilizes GPU parallel computing to accelerate computation by Monte Carlo. It incorporates a smoothing technique to not only fix the convergence issue in the Newton method but also improve the precision of estimates. Using examples, we illustrated that our novel algorithm outperforms the standard method in the real- world design of clinical trials. Our method not only boosts the computing speed to solve the design problem within a reasonable amount of time, but also improves the accuracy of estimations. We implement our algorithm using R language with Python functions called inside R. We will make our software available via GitHub and CRAN after the manuscript is accepted, so that all researchers can use to design their clinical trials.

This research is motivated by the problem of the design of clinical trials. Our software is developed to help the design of clinical trials. However, our method of using Monte Carlo, GPU computing, and smoothing can solve other general optimization problems with high-dimensional integration.	
	

	\section*{ACKNOWLEDGEMENTs}  The research was funded by the Natural Sciences and Engineering Research Council of Canada (NSERC) Discovery Grants (JZ and XZ) and Canada Research Chair Grant (YL and XZ). This research was enabled in part by support provided by WestGrid (www.westgrid.ca) and Compute Canada (www.computecanada.ca).

	\section*{CONFLICT OF INTEREST} The authors declare that the research was conducted in the absence of any commercial or financial relationships that could be construed as a potential conflict of interest.

	\section*{Appendix}
	\subsection*{A.1   Derivation of the Covariance Matrix of Normal Distribution in Formula~(\ref{Formula1})}\label{sec4.3}
	As discussed in \cite{ref2}, the correlation efficient of Z-statistics between entire population $X_1$ and the $k$-th subpopulation $X_k$ is 
	\begin{equation}\nonumber
	cor(X_1,X_k)= \sqrt{r_k}, \;\;\;\; \mbox{ for } k = 2,\ldots,n
	\end{equation}
	Consider any two (the $k$-th and the $l$-th) nested subpopulations for $2\leq k < l\leq n$. The $l$-th subpopulation is a subset of the $k$-th subpopulation and the proportion is $\frac{r_l}{r_k}$. We treat $X_k$ as the new `full' population, so we have
	\begin{equation}\nonumber
	cor(X_k, X_l) = \sqrt{\frac{r_l}{r_k}},  \;\;\;\; \mbox{ for }  2\leq k < l\leq n
	\end{equation}
	
	\subsection*{A.2 Derivation of the Upper bound of Variance for the Estimated Power}\label{sec4.2}
	Let 
	\begin{equation}\nonumber
	\hat{P}_{kl}= \delta(x^{(l)}_j \leq Z_{1-\alpha_j} - \sqrt{r_j I_3}\Delta^{(k)}_j, \forall j \in \{1,\ldots,n\})
	\end{equation}
	Then, we have, for $k = 1,\ldots, N_1, l= 1,\ldots, N_2$,
	\begin{equation}\nonumber
	\begin{aligned}
	&E(\hat{P}_{kl})= 1-p(\alpha)\\
	&V(\hat{P}_{kl})= p(\alpha)(1-p(\alpha))\leq\frac{1}{4}
	\end{aligned}
	\end{equation}
	From Formula~(\ref{Formula7}), we obtain
	\begin{equation}\nonumber
	Var(\hat{P}(\alpha))= \frac{1}{N_1^2N_2^2}\sum_{k=1}^{N_1}\sum_{l=1}^{N_2}Var(\hat{P}_{kl})\leq\frac{1}{4N_1N_2}
	\end{equation}

	\subsection*{A.3 Proof of the Smoothness property of P($\alpha_1,...,\alpha_{n-1}$)} \label{sec4.1}
	Without loss of generality, we prove $P(\balpha)$ is a smooth (infinitely differentialble) function of $\balpha$ for $n=1$. From Formula~(\ref{Formula4}), we have 
	\begin{equation}\nonumber
	\begin{aligned}
	P(\alpha) = 1-\int_{-\infty}^{\infty}\Phi_\Sigma(Z_{1-\alpha}-\sqrt{r_1 I_3}\Delta_1)
	\times f(\Delta_1)d\Delta_1
	\end{aligned}
	\end{equation}
	
	Then,
	\begin{equation}\nonumber
	\begin{aligned}
	\frac{dP(\alpha)}{d\alpha}&= -\int_{-\infty}^{\infty}\phi_\Sigma(Z_{1-\alpha}-\sqrt{r_1I_3}\Delta_1)\frac{dZ_{1-\alpha}}{d\alpha}f(\Delta_1)d\Delta_1\\
	&= \int_{-\infty}^{\infty}\phi_\Sigma(Z_{1-\alpha}-\sqrt{r_1I_3}\Delta_1) \frac{1}{\phi_\Sigma(Z_{1-\alpha})} f(\Delta_1)d\Delta_1
	\end{aligned}
	\end{equation}
	where $\phi_\Sigma(\cdot)$ is the density function of the stadard normal distribution. It is clear that $\frac{dP(\alpha)}{d\alpha}$ is a continuous function of $\alpha$ for all $0<\alpha<1$. Similarly, we could abtain the second, third and fourth derivation of $P(\alpha)$, which are all continous functions of $\alpha$. Thus, $P(\alpha)$ is a smooth function of $\alpha$, for all $0<\alpha<1$.

	\subsection*{A.4 The links of rotatable 3D plots corresponding to subfigures in Figure~\ref{Fig.1}} \label{sec4.4}
	
\textbf{No biomarker effect}:\\
Fitted TPS Surface of the optimal power : http://www.math.uvic.ca/\textasciitilde xuekui/HDDesign/Figure1.2 \\
Fitted TPS Surface of the optimal $\balpha$ : http://www.math.uvic.ca/\textasciitilde xuekui/HDDesign/Figure1.1

\noindent \textbf{Weak biomarker effect}:\\
Fitted TPS Surface of the optimal power : http://www.math.uvic.ca/\textasciitilde xuekui/HDDesign/Figure1.4 \\
Fitted TPS Surface of the optimal $\balpha$ : http://www.math.uvic.ca/\textasciitilde xuekui/HDDesign/Figure1.3

\noindent \textbf{Strong biomarker effect}:\\
Fitted TPS Surface of the optimal power : http://www.math.uvic.ca/\textasciitilde xuekui/HDDesign/Figure1.6 \\
Fitted TPS Surface of the optimal $\balpha$ : http://www.math.uvic.ca/\textasciitilde xuekui/HDDesign/Figure1.5

	\bibliographystyle{plain}
	\bibliography{HD_optimization}

\begin{thebibliography}{10}

\bibitem{ref18}
RH~Byrd, PH~Lu, J~Nocedal, and C~Zhu.
\newblock A limited memory algorithm for bound constrained optimization.
\newblock {\em SIAM J. Sci. Comput}, 16(5):1190--1208, 1995.

\bibitem{ref2}
C~Chen and RA~Beckman.
\newblock Hypothesis testing in a confirmatory phase iii trial with a possible
  subset effect.
\newblock {\em Statistics in Biopharmaceutical Research}, 1(4):431--440, 2009.

\bibitem{ref1}
C~Chen and RA~Beckman.
\newblock Efficient, adaptive clinical validation of predictive biomarkers in
  cancer therapeutic development.
\newblock {\em Advances in Experimental Medicine and Biology}, 867:81--90,
  2015.

\bibitem{ref8}
C~Chen, Nicole Li, Y~Shentu, L~Pang, and RA~Beckman.
\newblock Adaptive informational design of confirmatory phase iii trials with
  an uncertain biomarker effect to improve the probability of success.
\newblock {\em Statistics in Biopharmaceutical Research}, 8(3):237--247, 2016.

\bibitem{ref3}
MA~Cobleigh, CL~Vogel, and D~Tripathy.
\newblock Multinational study of the efficacy and safety of humanized anti-her2
  monoclonal antibody in women who have her2-overexpressing metastatic breast
  cancer that has progressed after chemotherapy for metastatic disease.
\newblock {\em Journal of Clinical Oncology}, 17(9):2639--2648, 1999.

\bibitem{ref7}
B~Freidlin and R~Simon.
\newblock Adaptive signature design: An adaptive clinical trial design for
  generating and prospectively testing a gene expression signature for
  sensitive patients.
\newblock {\em Clinical Cancer Research}, 11(21):7872--7878, 2005.

\bibitem{ref0}
EB~Garon, NA~Rizvi, and R~Hui.
\newblock Pembrolizumab for the treatment of non-small-cell lung cancer.
\newblock {\em New England Journal of Medicine}, 372(21):2018--2028, 2015.

\bibitem{ref21}
PJ~Green and BW~Silverman.
\newblock {\em Nonparametric Regression and Generalized Linear Models: A
  roughness penalty approach}.
\newblock Chapman and Hall/CRC, 1993.

\bibitem{ref14}
M~Hernández, GD~Guerrero, and JM~Cecilia.
\newblock Accelerating fibre orientation estimation from diffusion weighted
  magnetic resonance imaging using gpus.
\newblock {\em PLOS ONE}, 8(4):e61892, 2013.

\bibitem{ref16}
Christopher Jennison and BW~Turnbull.
\newblock {\em Group Sequential Methods with Applications to Clinical Trials}.
\newblock Chapman and Hall/CRC, 1999.

\bibitem{ref6}
W~Jiang, B~Freidlin, and R~Simon.
\newblock Biomarker-adaptive threshold design: A procedure for evaluating
  treatment with possible biomarker-defined subset effect.
\newblock {\em Journal of National Cancer Institute}, 99(13):1036--1043, 2007.

\bibitem{ref15}
P~Klus, S~Lam, and D~Lyberg.
\newblock Barracuda-a fast short read sequence aligner using graphics
  processing units.
\newblock {\em BMC Res Notes 5}, 5(1), 2012.

\bibitem{ref5}
TJ~Lynch, DW~Bell, R~Sordella, and S~Gurubhagavatula.
\newblock Activating mutations in the epidermal growth factor receptor
  underlying responsiveness of non-smallcell lung cancer to gefitinib.
\newblock {\em New England Journal Medicine}, 350(21):2129--2139, 2004.

\bibitem{ref22}
D~Nychka, R~Furrer, J~Paige, and S~Sain.
\newblock fields: Tools for spatial data, 2017.
\newblock R package version 10.0.

\bibitem{ref12}
T~Ohl.
\newblock Vegas revisited: Adaptive monte carlo integration beyong
  factorization.
\newblock {\em Computer Physics Communications}, 120(1):13--19, 1999.

\bibitem{ref13}
Kawabata Setsuya.
\newblock A new version of the multi-dimensional integration and event
  generation package bases/spring.
\newblock {\em Computer Physics Communications}, 88:309--326, 1995.

\bibitem{ref4}
DJ~Slamon, B~Leyland-Jones, and S~Shak.
\newblock Use of chemotherapy plus a monoclonal antibody against her2 for
  metastatic breast cancer that overexpresses her2.
\newblock {\em New England Journal of Medicine}, 344(11):783--792, 2001.

\bibitem{ref11}
Hong-Zhong Wu, JunJie Zhang, Long-Gang Pang, and Qun Wang.
\newblock Zmcintegral: A package for multi-dimensional monte carlo integration
  on multi-gpus.
\newblock {\em Computer Physics Communications}, 248:106962, 2020.

\bibitem{ref20}
C~Zhu, RH~Byrd, Peihuang Lu, and Jorge Nocedal.
\newblock L-bfgs-b: Algorithm 778: L-bfgs-b, fortran routines for large scale
  bound constrained optimization.
\newblock {\em ACM Transactions on Mathematical Software}, 23(4):550--560,
  1997.

\end{thebibliography}
	
\end{document}